# On the Rayleigh theorem for inflectional velocity instability of inviscid flows

Hua-Shu Dou

Temasek Laboratories

National University of Singapore,

Singapore 117508

Email: tsldh@nus.edu.sg; huashudou@yahoo.com

**Abstract**: It is exactly proved that the classical Rayleigh Theorem on inflectional velocity instability is wrong which states that the necessary condition for instability of inviscid flow is the existence of an inflection point on the velocity profile. It is shown that the disturbance amplified in 2D inviscid flows is necessarily 3D. After the break down of T-S wave in 2D parallel flows, the disturbance becomes a type of spiral waves which proceed along the streamwise direction. This is just the origin of formation of streamwise vortices.

**Keywords**:  Rayleigh theorem; Flow instability; Inviscid flow; Inflection.

PACS numbers:
47.10.+g        General theory in fluid dynamics
47.15.Fe        Stability of laminar flows
47.20.Ft        Instability of shear flows
47.27.Cn        Transition to turbulence
47.32.cd        Vortex stability and breakdown

## 1. Introduction

In the classical theory for flow instability, Rayleigh (1880) first developed a general linear stability theory for inviscid parallel shear flows, and showed that a necessary condition for instability is that the velocity profile has a point of inflection [1]. Heisenberg (1924) showed that if a velocity distribution allows an inviscid neutral disturbance with finite wave-length and non-vanishing phase velocity, the disturbance with the same wave-length is unstable in the real fluid when the Reynolds number is sufficiently large [2]. Later, Tollmien (1935) succeeded in showing that Rayleigh's criterion also constitutes a sufficient condition for the amplification of disturbances for velocity distributions of the symmetrical type or of the boundary-layer type [3]. Lin (1944) mentioned that he has shown where a point of inflexion exists in the velocity curve, but a neutral disturbance does not exist [4]. About the dual roles of viscosity, he was able to demonstrate the different influences of viscosity on the disturbance amplification with low Re and high Re. His conclusions are as follows. For small viscosity, the effect of viscosity is essentially destabilizing and an increase of Re gives more stability. For large viscosity (low Re), viscosity plays a stable role by dissipation of energy. Fjϕrtoft (1950) gave a further necessary condition for inviscid instability, that there is a maximum of vorticity for instability [5]. Therefore, it is well known that inviscid flow with inflectional velocity profile is unstable, while inviscid flow with no inflectional velocity profile is stable [6-10]. The associated analysis showed that the effect of viscosity is complex, and it plays dual roles to the flow instability. The linear stability theory has been confirmed by the famous experiment by Schubauer and Skramstad [11], which proved that there exists really a 2D wave under low noised environment and this wave was named as Tollmien-Schlichting wave [6-10].

Recently, we proposed a new theory, named as energy gradient theory, to explain the flow instability and transition to turbulence [12-14]. The critical condition calculated at turbulent transition determined by experiments obtains consistent agreement with the available experimental data for parallel flows and Taylor-Couette flows [15]. When the theory is considered for both parallel and curved shear flows, three important theorems have been deduced. These theorems are: (1)Potential flow (inviscid and irrotational) is stable. (2) Inviscid rotational (inviscid and nonzero vorticity) flow is unstable. (3) Velocity profile with an inflectional point is unstable when there is no work input or output to the system, for both inviscid and viscous flows. From the theorem (3), it is demonstrated that the existence of an inflection point on velocity profile is a sufficient condition, but not a necessary condition for flow instability, for both inviscid and viscous flows. Following these results, it is presumed that the classical Rayleigh theorem is wrong which states that a necessary condition for inviscid flow instability is the existence of an inflection point on the velocity profile. In present study, we show rigorously the proof why Rayleigh theorem is wrong. This can be demonstrated with the governing equation of vorticity and the Helmholtz vortex theorem. Then, two new theorems are deduced.

## 2. Rayleigh Equation

Let the mean flow, which may be regarded as steady, be described by its Cartesian velocity components U,V,W and its pressure P, the corresponding quantities for the non-steady disturbance will be denoted by u, v, w (u in streamwise, v in transverse, and w in spanwise directions), and p, respectively. Hence, in the resultant motion the velocity components and the pressure are

$$u = U + u', \; v = V + v', \; w = W + w', \; p = P + p'. \qquad (1)$$

Substituting above expressions into the Euler equation for inviscid flow and substracting the equation for base flow, the linearized equation of disturbance can be obtained [1,6-10].

It is assumed that the disturbance is two-dimensional (2D), then a stream function is introduced. The stream function representing a single oscillation of the disturbance is assumed to be of the form

$$\psi(x, y, t) = \phi(y)e^{i(\alpha x - \beta t)}, \quad (2)$$

where $\alpha$ is a real quantity and $\beta$ is a complex quantity, $\beta = \beta_r + i\beta_i$. Dividing $\beta$ by $\alpha$, a complex quantity c is obtained, $c = \beta/\alpha = c_r + ic_i$, here, $c_r$ is the speed of the wave propagating and $c_i$ expresses the degree of damping or amplification of the disturbance ($c_i$ =0, neutral disturbance; $c_i$ <0, the disturbance decays; $c_i$ >0, the disturbance amplified). Thus,

$$u' = \frac{\partial \psi}{\partial y} = \phi'(y)e^{i(\alpha x - \beta t)}, \quad (3)$$

$$v' = -\frac{\partial \psi}{\partial x} = -i\alpha\phi(y)e^{i(\alpha x - \beta t)}. \quad (4)$$

Introducing these values into the linearized equation of the disturbance, the following ordinary differential equation is obtained [1,6-10],

$$(U - c)(\phi'' - \alpha^2 \phi) - U''\phi = 0, \quad (5)$$

which is known as the frictionless stability equation, or Rayleigh's equation.

## 3. Rayleigh's necessary condition for instability of inviscid flows

Let's multiply Eq.(5) by its complex conjugate, we obtain [1, 6-10]

$$\phi^* \phi' - \alpha^2 \phi \phi^* - \frac{U''}{U - c}\phi\phi^* = 0. \quad (6)$$

Then, integrating above equation by part over y, the imaginary part of the resulting equation is

$$c_i \int_{y_1}^{y_2} \frac{U''|v'|^2}{|U - c|^2} dy = 0. \quad (7)$$

If the disturbance is amplified, $c_i$ is larger than zero. It can be seen that for the equality to be valid $U''$ has to change sign over the integration space. Thus, there should be at least one point over the distance between $y_1$ and $y_2$ at which $U''$ =0. In other words, it is necessary that there is an inflection point on the velocity profile.

## 4. Three-dimensionality of the disturbance amplification

Equation (7) is derived under the assumption of disturbance to be 2D. Thus, equation (7) is only correct when the disturbance is 2D. However, we will prove in the following when the disturbance is amplified ($c_i$ >0), the disturbance must become three-dimensional (3D). In other words, $c_i$ >0 should correspond to $w' \neq 0$. At such condition, the integration in Eq.(7) should use the information of 3D disturbance (including spanwise disturbance). Therefore, Eq.(7) is not established anymore for $c_i$ >0 since it uses only the 2D information for the integration. As a result, one could not obtain the Rayleigh theorem because Eq.(7) is not established for 3D disturbance.

Now, we begin to prove the three-dimensionality of the disturbance after instability sets in using the Helmholtz vortex theorem. The governing equation for vortex transportation can be written as

$$\frac{D\omega}{Dt} \equiv \frac{\partial \omega}{\partial t} + V \cdot \nabla \omega = \nu \nabla^2 \omega + V \times \nabla \times \omega . \qquad (8)$$

For inviscid 2D flows,

$$\frac{D\omega}{Dt} \equiv \frac{\partial \omega}{\partial t} + V \cdot \nabla \omega = 0 . \qquad (9)$$

It is observed from Eq.(9) that the vorticity in 2D inviscid flows can not be changed. This is the famous Helmholtz vortex theorem. When the disturbance is amplified, the disturbance wave u' and v' will be changed. Thus the vorticity $\omega$ will also be changed, which is not possible in 2D inviscid flows. Therefore, it is necessary that the disturbance becomes 3D if it is amplified. This means that stretch/compression of vortex is necessary. We obtain the following theorem:

**Theorem (1): The disturbance amplified in 2D inviscid flows is necessarily 3D.**

From above discussions, the meaning of neutral stability curve ($c_i$ =0) for Raleigh equation (the viscous counterpart is Orr-Somerfeld equation) is the boundary between 2D and 3D disturbances (Fig.1). In other words, it is the critical condition for 3D disturbance initiation.

When a disturbance is 2D and propagates along the streamwise direction, the disturbance velocity u' and v' are within a plane (Fig.2). This is the case of Tollmien-Schlichting waves. From theorem (1), the instability of (2D) Tollmien-Schlichting waves generates 3D disturbances in which there is a phase difference between the velocity disturbances. Thus, the 2D plane waves become spiral waves in streamwise direction after instability sets in (Fig.3).

**Theorem (2): After the breakdown of Tollmien-Schlichting waves in 2D parallel flows, the disturbance becomes a type of spiral waves which proceed along the streamwise direction.**

These spiral waves are equivalent to the Taylor vortices generated between concentric rotating cylinders. They both are the products of linear instability, and the base flows are still laminar flows after the instability sets in. These vortices are nothing to do with turbulence if no further development occurs (i.e., increases in Re and disturbance amplitude).

In the text books [6-10], it is shown that an oblique disturbance propagating along a direction making an angle with x direction can be transformed into psudo-2D disturbance via Squire transformation [16]. In such a way, a streamwise function still can be used to obtain the Rayleigh equation. However, this type of oblique wave is not arbitrary 3D, but a plane wave in which all the disturbance are within a plane along a direction which makes an angle with x axis. The disturbance in the direction normal to this plane is zero, and the disturbance u' and w' has same phase in this wave.

For arbitrary 3D disturbance, the disturbance could not be limited in a plane due to phase difference among u', v', and w'. In this case, u', v' and w' form a spiral wave and the vector of disturbance follows a spiral trace. The Squire's transformation is invalid for this case. Thus, a stream function does not exist. Therefore, the Rayleigh equation (the viscous counterpart is the Orr-Sommerfeld equation) can not be established in this case.

From theorem (1), it is known that the instability of 2D laminar flows with T-S waves can only become 3D laminar flows, and it is impossible for 2D laminar flow directly to transit to 2D turbulent flow. The turbulence must be 3D, and there is no turbulence in 2D flows [17]. An extension of a fluid particle in one direction from an equilibrium state must cause the compression in the other two directions, and vice versa.

Experimental data showed that in the transition to turbulence in boundary layer flows, under small disturbance environment, some staggered or unstaggered "lambda" shaped pattern could be generated; while under larger disturbance, streamwise vortices could be generated [7-8]. These "lambda" shaped patterns can be the indication of 3D disturbance development after linear instability of 2D Tollmien-Schlichting waves sets in. When the disturbance becomes large, the spiral waves are strong, and they would form streamwise vortices via vortices merging process. Therefore, it is concluded that the instability of 2D laminar flow necessarily lead to 3D flows. Linear instability of 2D laminar flow leads to 3D flow but to be laminar flow. Only nonlinear instability of 3D laminar flow could result in turbulence; the base flow before nonlinear instability is already 3D. The spiral behavior of the propagation of traveling waves is the key of generation of the streamwise vortex and hairpin vortex and the events in the transition.

**5. Conclusions**

It is proved that the disturbance amplified in 2D inviscid flows is necessarily 3D. The classical Rayleigh theorem on inflectional velocity instability of inviscid flows is wrong since it does not account for the 3D of the disturbance for $c_i$ >0. After the breakdown of Tollmien-Schlichting waves in 2D parallel flows, the disturbance becomes a type of spiral waves which proceed along the streamwise direction. The spiral behavior of the propagation of traveling waves is the origin of the streamwise vortex and hairpin vortex and the events in the transition.

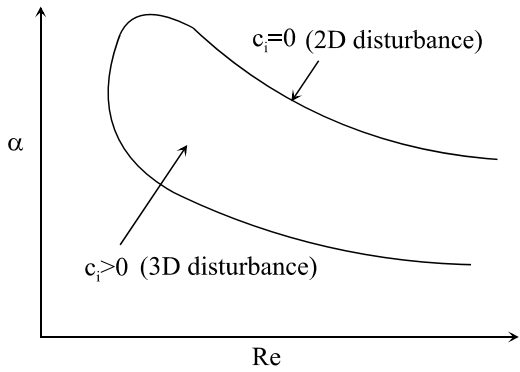

Fig.1 Physical meaning of neutral stability curve for boundary layer flow.
Re: Reynolds number; α: wave number.

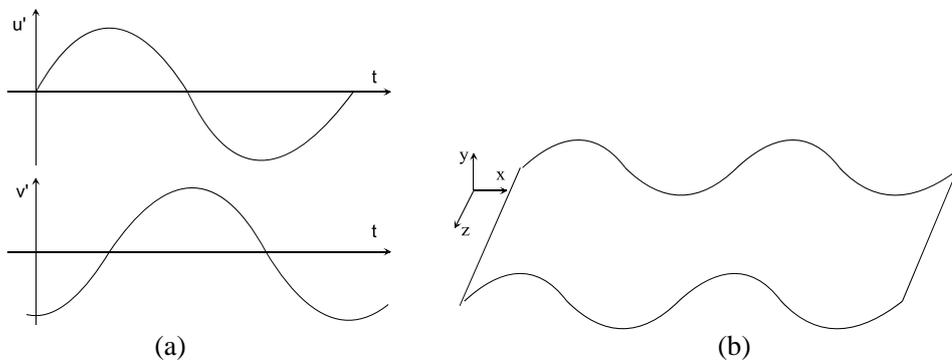

(a)                               (b)

Fig.2 (a) Two-dimensional disturbance; (b) Form of wave. Disturbance velocity u' and v' are within a plane. This is the case of Tollmien-Schlichting waves.

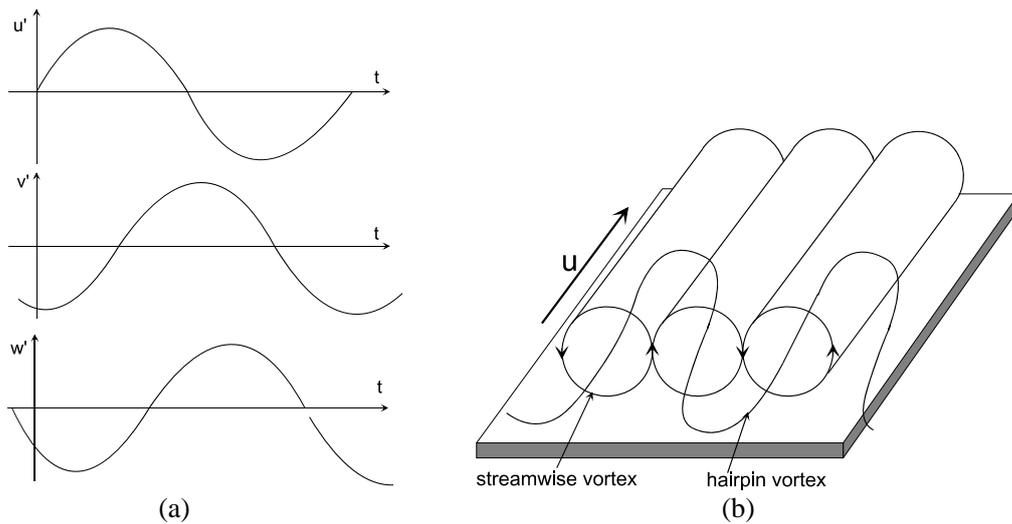

(a)                               (b)

Fig.3 (a) Three-dimensional disturbance; (b) Streamwise vortices formation. Disturbance velocity u', v' and w' are not within a plane. There are phase differences between the velocity components. Thus, a spiral wave is formed. This is the origin of streamwise vortices.